\documentclass[12pt]{iopart}

\usepackage{color}
\usepackage{times}
\usepackage{fancyhdr}
\usepackage{float}
\usepackage{alltt}
\usepackage{listings}
\usepackage{enumerate}
\usepackage[T1]{fontenc}
\usepackage{ae}
\usepackage{rotating}
\usepackage{subfigure}
\usepackage{amssymb}
\usepackage{graphicx}
\usepackage{dcolumn}
\usepackage{bm}
\usepackage{aas_makros_pk}

\def\be{\begin{equation}}
\def\ee{\end{equation}}
\def\beg{\begin{align}}
\def\eeg{\end{align}}
\def\bi{\begin{itemize}}
\def\ei{\end{itemize}}
\def\ben{\begin{enumerate}[1.]}
\def\een{\end{enumerate}}

\newcommand{\bo}{\raise-1mm\hbox{\Large$\Box$}}

\newcommand{\hrss}{h_{\mathrm{rss}}}

\newcommand{\egw}{E_{\mathrm{GW}}}

\newcommand\ligodoc{P1300018}

\begin{document} 

\title[Excluding Source Models with Multiple Astrophysical Observations]{Excluding Source Models with Multiple Astrophysical Observations}

\newcommand*{\CH}{California Institute of Technology, 1200 E California Blvd., Pasadena, CA 91125, USA}
\author{Peter~Kalmus}  
\address{\CH}
\ead{peter.kalmus@ligo.org}
\author{Michele~Zanolin}
\address{Embry Riddle University, 3700 Willow Creek Road, Prescott Arizona, 86301, USA}
\author{Sergey~Klimenko}
\address{University of  Florida, P.O. Box 118440, Gainesville, Florida, 32611, USA}

\begin{abstract}

We describe a general method to observationally exclude a theoretical model for gravitational wave (GW) emission from a transient astrophysical source (event) by using a null detection from a network of GW detectors. In the case of multiple astrophysical events with no GW detection, statements about individual events can be combined to increase the exclusion confidence. We frame and demonstrate the method using a population of hypothetical core collapse supernovae.

\end{abstract}
\maketitle


\section{Introduction}

The global network of gravitational wave (GW) detectors\,\cite{detectorLivingReview} has yet to discover a signal, but their null detections have already constrained astrophysical models.  Magnetar flares provide an example of such constraints. Satellite detectors observe unpredictable bursts of soft gamma-rays from neutron stars with extreme magnetic fields, known as magnetars. The hypothesis that a GW arrives within $\pm 2$\,s of such a gamma-ray burst was tested by looking for transient excess power in the GW data within this signal region, and comparing to the background\,\cite{a5sgr}.  There was no GW detection. The loudest transient event in the signal region was then compared to simulated GW signals predicted by models of the damping of non-radial global stellar modes, allowing the null detections to constrain these models.

So long as the signal region duration $t_s$ is much less than the characteristic detector uptime timescale $t_d$, model exclusion from individual null detections can proceed as described above.  However, when $t_s \gg t_d$ the situation is more complicated.  In this case, the detectors may have outages during the signal region, and the data will have gaps. A short burst of signal falling a gap will be missed.

Electromagnetically-triggered searches for GWs from core-collapse supernovae (CCSNe) are an example.  A typical core collapse time estimated from optical light curves has a large uncertainty, and GW signal regions typically have durations lasting days or weeks.  For example, SN2007gr was a promising supernova about 9\,Mpc away. It was discovered on August 15, 2007, and the a pre-discovery empty image from August 10 constrains the core-collapse time to an approximately five day region. On the other hand, the typical GW detector uptime timescale is only hours. Because $t_s \gg t_d$, using a null detection from GW data to exclude SN models with a single CCSN event will be complicated by gaps in the data, and exclusion statements must be made with lowered confidence to reflect these observational gaps.  This motivates the use of multiple CCSN events to improve the confidence of exclusion statements.  

In this paper we describe a simple and general method for quantifying the confidence of null-detection model exclusion statements given non-stationary detectors (or detector networks) and multiple observational events.  We illustrate the method with a hypothetical search for GWs from CCSNe.

\section{Method}

\subsection{Loudest event limits} \label{sec:loudestEventLimits}

We begin by reviewing the frequentist loudest event limit construction, consistent with procedures described in\,\cite{feldman98} and used in astrophysical GW searches for burst-like signals such as\,\cite{S2S3S4GRB, s5y1sgr, stackSgr, a5sgr}. Any physical observable which can be expressed as a monotonic function of the signal amplitude may be constrained; we refer to it as the observable of interest.  In the case of GW searches, the observable of interest is typically either root-sum-square GW strain at the detector ($\hrss$) , GW emission energy $\egw$, or source distance ($d \propto \hrss^{-1}$). The signal is convoluted with the detector's response function and noises and may produce an ``analysis event.''   In general an analysis event is anything interesting in the analyzed data, and could be a time- and frequency-limited burst of excess power, for example\,\cite{kalmus07}.  An analysis event in the signal region might correspond to either noise or a transient GW signal that is not strong enough to claim a detection.   The analysis must define a ranking statistic with which to calculate the `loudness' of analysis events.  

The general procedure for constructing an upper or lower limit is as follows:
\begin{enumerate}[(1)]
\itemsep 1pt
\parskip 0pt
 \item Using the ranking statistic, estimate the loudness of the analysis events in the signal region. 
 \item For a class of signals predicted by the model of interest,  add a simulated signal at a particular amplitude at a randomly selected time into the data.  Analyze the data. If the chosen amplitude is large enough, the simulated signal will produce an analysis event with some measurable loudness.  \label{item:inject}
 \item Assign an analysis event to the simulated signal. To do this, find analysis events within a time $\Delta t$ of the simulated signal's known placement, estimate their loudnesses, rank them, and choose the loudest.  $\Delta t$ is chosen to be as small as possible while not missing a significant fraction of loud, easily detectable signals\,\cite{kalmus08}.   The loudness of the analysis event corresponding to the recovered simulated signal will depend on local noise fluctuations, in addition to the amplitude of the simulated signal. \label{item:detect}
 \item Repeat (\ref{item:inject}) and (\ref{item:detect}) for a range of signal amplitudes (which can be mapped to the physical observable of interest $x$) and at each value determine the fraction $P(x)$ of  added simulated signals with associated analysis events louder than the loudest signal region analysis event.    \label{item:range}
 \item Repeat (\ref{item:range}) using different simulated signal types representing different models as necessary.  
\end{enumerate}

\begin{figure}[!t]
\begin{center}
\includegraphics[angle=0,width=115mm,  clip=true]{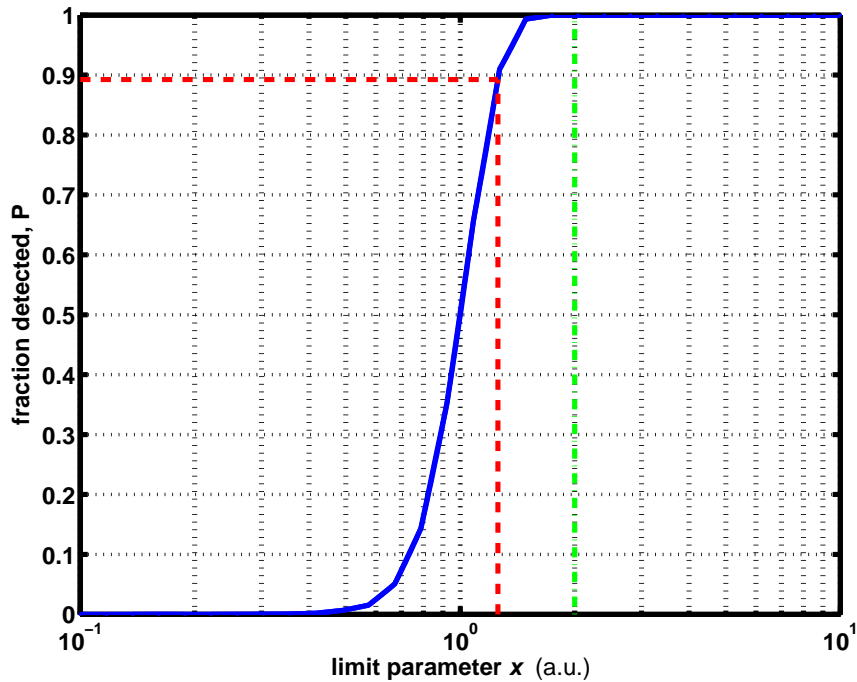}
\caption{ Example efficiency curve for a hypothetical null detection of an observational event, illustrating a model exclusion for a model predicting an observable of interest $x_M=2$ (green dash-dot line) and the construction for a 90\% detection efficiency loudest event limit (red dashed lines, in this case, an upper limit) described in the text.  This efficiency curve was constructed using a simple Monte Carlo simulating detector noise.  In the absence of detector noise, the efficiency curve would be a step function.  } \label{fig:effCurve}
\end{center}
\end{figure}

For a single signal region with 100\% duty cycle,  this procedure allows us to construct efficiency curves $P(x)$ which give the probability that the signal producing a particular measurement of the observable of interest $x$ will be louder than the loudest signal region event.  An example is shown in Figure\,\ref{fig:effCurve}.   

The conclusion we draw from $P(x)$ depends on a particular model $M$.  If $M$ predicts the value of the observable of interest associated with the observational event, $x_M$ , then we can state the exclusion confidence $P(x_M)$.  This is shown by the vertical green line in Figure\,\ref{fig:effCurve}.

If $M$ predicts a range in $x$, we may be able to exclude a region of model-space.  We can find the 90\% detection efficiency loudest event upper limit $x^{90\%}$, which is the value of $x$ at which 90\% of analysis events associated with simulations predicted by $M$ would be louder than the loudest signal region event.  This results in the 90\% confidence limit for the model $M$ and the observational event, the red construction in Figure\,\ref{fig:effCurve}.  We see that loud analysis events in the signal region diminish the excluding power of the data.

\subsection{Non-stationary detector observing one event}

We now discuss the case of one observational event with a model-predicted signal region covered by a non-stationary detector.  A detector is non-stationary if its properties change over time.  Here, we consider changes in detector sensitivity, or the detector turning on and off.  A network of multiple detectors independently turning on and off can be thought of as a single non-stationary detector comprised of the individual instruments.  

\begin{figure}[!t]
\begin{center}
\includegraphics[angle=0,width=115mm,  clip=true]{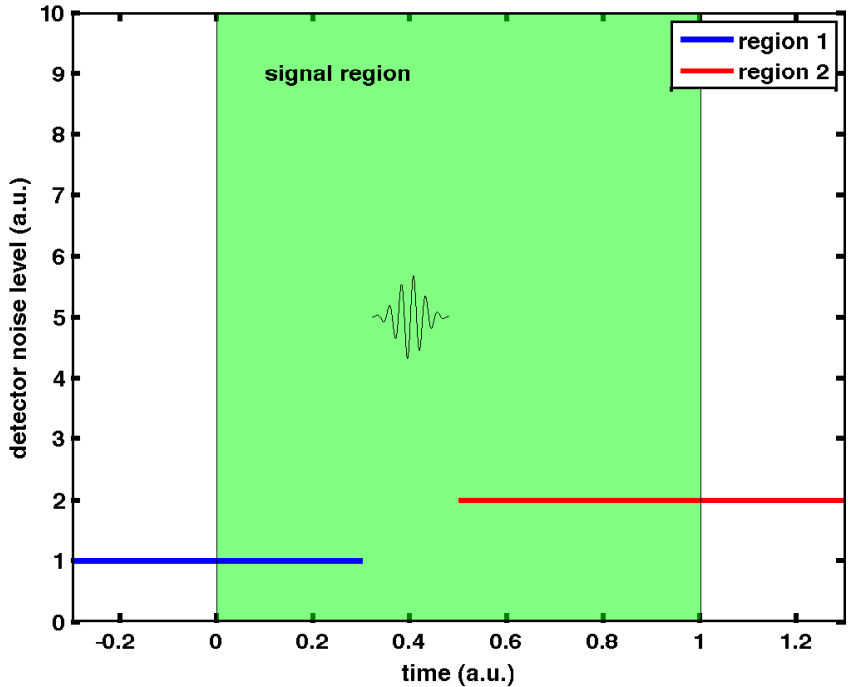}
\caption{ Example of a signal region (shaded) covered by multiple stationary detectors.  Region 1 (the blue line) has a duration of the first 30\% of the signal region and is covered by a detector or detector network observing with a noise level of 1.  Region 2 (the red line) has a duration of the last 50\% of the signal region and is covered by a detector or detector network observing with a noise level of 2.  Here a lower noise level corresponds to a higher sensitivity, as with GW strain (thus the detector covering region 2 is half as sensitive as the detector covering region 1).  The overall duty cycle is 80\%.  We cannot exclude any model which predicts this signal region, as a signal may have occurred while no detector was observing. } \label{fig:multinetworkSetup}
\end{center}
\end{figure}

An example of this scenario is shown in Figure\,\ref{fig:multinetworkSetup}; the signal region predicted by the model is shaded.  Here, region 1 (the blue line) has a duration of the first 30\% of the signal region. It is covered by a detector (or detector network) observing with a noise level of 1, which turns off at $t=0.3$.  In Region 2 (the red line), the detector turns back on at $t=0.5$ but is now noisier, with a noise level of 2.  Between these regions, 20\% of the signal region occurs with no detector observing; the duty cycle over the signal region is 0.8.  A signal may have occurred while no detector was observing, so we cannot use this event to absolutely exclude the model.

\begin{figure}[!t]
\begin{center}
\includegraphics[angle=0,width=115mm,  clip=true]{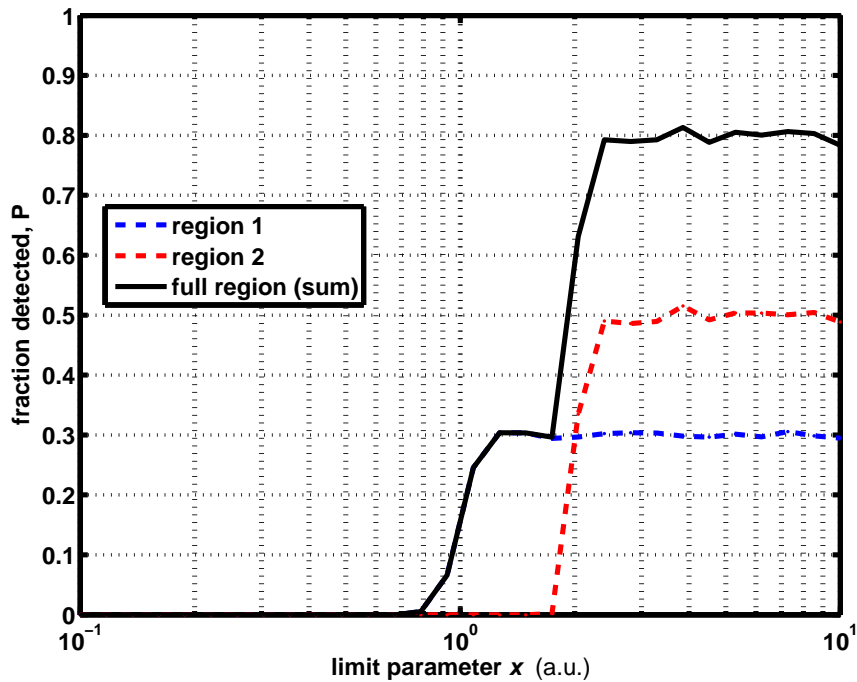}
\caption{ Monte Carlo demonstration of efficiency curves corresponding to the example signal region shown in Figure\,\ref{fig:multinetworkSetup}.  The black curve shows the efficiency curve for the total signal region; it achieves a maximum value 80\% corresponding to the duty cycle of the signal region.  The blue and red dashed curves are efficiency curves constructed in the blue and red portions of the signal region, respectively. The sum of these two curves equals the black curve. In region 1 the detector was more sensitive, and is able to sense signals corresponding to smaller values of $x$. } \label{fig:multinetworkResult}
\end{center}
\end{figure}

This case can be analyzed in the same way as the simpler single detector, 100\% duty cycle case introduced above:  add simulated signals randomly into the entire signal region and construct the efficiency curve. The procedure requires no modification. Simulated signals which fall into regions where no detector is observing will, of course, not be detected, and will not be counted in the detected fraction of the efficiency curve. The resulting efficiency curve is shown in Figure\,\ref{fig:multinetworkResult}.  While in this example the detectors change discretely, the same procedure will also work in the case of continuously changing detectors. (In practice, GW detectors change continuously.)

We assume a flat probability density function for the location of the observational event (for example, a CCSN) within this signal region.  We can make an exclusion statement for a model $M$, which predicts a signal from the event with some specific value $x_M$ for the observable of interest  (which could correspond to the CCSN GW strain, for example) following the procedure in Section\,\ref{sec:loudestEventLimits}, except in this case the efficiency curve will asymptote to the overal duty cycle $P(x=\infty)=0.8$.    We could make a statement such as,``We exclude the model $M$ with 80\% exclusion confidence'' if $x_M$ were greater than three, for example.   With a single event this is the best we can do.  For any value of $x_M$ predicted by the model, we can find the corresponding fraction detected, which can be interpreted as the exclusion confidence.

We note that the method can be generalized to the case where the probability density function for the event location with the signal region $t_s$ is assigned a non-flat prior.  In this case the time location for each simulated signal within the signal region is chosen randomly according to the prior PDF, whatever that may be.   No other change is necessary.

\subsection{Multiple events}

Consider two independent but identical astrophysical events, occurring within distinct signal regions each with detector duty cycle of 0.5, neither of which is detected.  The probability of missing detectable signals from both events is 1/4, and we can exclude the model $M$ with up to 75\% exclusion confidence.  In general, for $N$ independent observational events with duty cycles $C_i$ (the fraction of the signal region observed by the detector for the $i$th event), the maximum exclusion confidence is
 \be
 	\epsilon_{\mathrm{max}} = 1-\prod_{i=1}^{N} (1-C_i).
 \ee

It is possible that some of the null detections from these observational events will not fully constrain the model, given the detector sensitivity.  In this case, we  construct the individual efficiency curves and find the values $P_i(x_M)$ for each observational event.  Then the observational exclusion confidence is
 \be
 	\epsilon = 1-\prod_{i=1}^{N} (1- P_i(x_M)).
 \ee

For example, consider a CCSN model $M_{\mathrm{SN}}$ which predicts an energy of GW emission $\egw$ and a particular GW emission pattern during the CCSN event.  We can constrain the model using observations from multiple CCSN events at known distances $d_i$.  

First,  efficiency curves are constructed for the individual events, $P_i (d)$.  Here, the observable of interest is source distance, $d$, and the amplitudes of the simulations predicted by $M_{\mathrm{SN}}$ are scaled according to the inverse of the hypothetical distance to the CCSN (we imagine letting the source distance vary).  Then the known distance to the source $d_i$ is drawn as a vertical line on the curve and the model exclusion confidence $P_i(d_i)$ is read off from the y-axis.

These $P_i(d_i)$  are combined into the overall exclusion confidence, $\epsilon$:
\be
 \epsilon = 1 - \prod_{i =1}^N (1-P_i(d_i)) \label{eq:reach}
\ee

To illustrate, we now construct a specific example involving GWs from CCSNe.  The emitted energy in gravitational waves $\egw$ from some GW signal is proportional to the square of the signal's root-sum-square-strain $\hrss^2$ and the square of the distance to the source $d$ (by conservation of energy):
\be
\egw \propto d^2 \hrss^2.
\ee

In units where the proportionality becomes equality, consider three CCSN events, at distances of 1, 2 and 4 units with signal region duty cycles of 0.4, 0.5 and 0.6 respectively.  Imagine a detector and loudest event combination which always yields a detection when the simulated signal $\hrss$ is $\ge 1$, and a model that emits nine units of GW energy which yields $\hrss = 1$ at a distance of 3 units.  

\begin{figure}[!t]
\begin{center}
\includegraphics[angle=0,width=115mm,  clip=true]{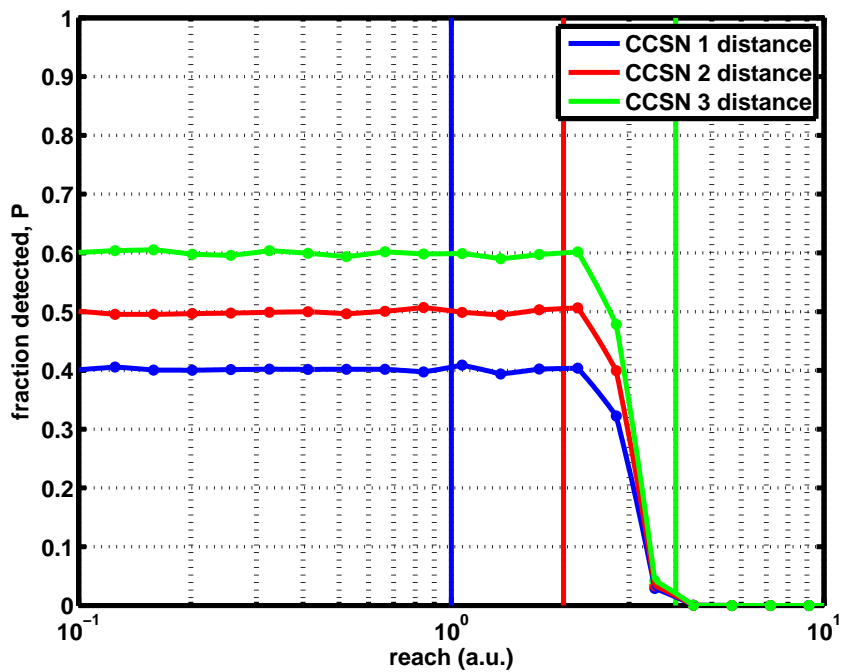}
\caption{ Distance efficiency curves for three hypothetical CCSN events.  The blue and red vertical lines at distances of 1 and 2 units cross their respective curves at detection probabilities of 0.4 and 0.5 respectively, while CCSN 3 at a distance of 4 units is too far away to be detected.  Equation\,\ref{eq:reach} yields an exclusion confidence of 0.7. } \label{fig:reachApproach}
\end{center}
\end{figure}

The individual efficiency curves are constructed for each CCSN in Figure\,\ref{fig:reachApproach}.  For illustrative purposes we have added some randomized uncertainty in each simulated signal's $\hrss$ to model detector noise, which is why the efficiency curves are not perfect step functions.   The detector can detect GW signals out to a distance of 3 units, although in some cases noise fluctuations will subtract from the signal while in others they will add to the signal.  Given the noise level used here, the CCSN events at distances of 1 and 2 units should be detected, and using Equation\,\ref{eq:reach} the exclusion confidence for the model given the three CCSN events is 0.7.  For simplicity, our example avoids the transition regions of the efficiency curves. However, Eq.\,\ref{eq:reach} will work regardless of where the known distances fall within the efficiency curves. 

We are free to choose a different observable of interest for the same physical processes.  In the example above we could equally well begin by constructing efficiency curves  for the individual observational events using emitted GW energy, $\egw$.  These individual efficiency curves $P_i (\egw)$  are combined into the total exclusion confidence curve, $P_(\egw)$ via
\be
 P_T(\egw) = 1 - \prod_{i =1}^N [1-P_i (\egw)]
\ee

$P_T(\egw)$ gives the exclusion confidence as a probability as a function of $\egw$, which folds in the known distances $d_i$:
\be \egw = 4\pi
d_i^2 \frac{c^3}{16 \pi G} \int_{-\infty}^{\infty}\left(
(\dot{h}_{i +})^2 + (\dot{h}_{i \times})^2\right) dt. 
\ee
Then the energy predicted by the model $M$, $\egw^M$, is drawn as a vertical line on the curve and the total model exclusion confidence $\epsilon$ is read off from the y-axis.

\begin{figure}[!t]
\begin{center}
\subfigure[]{
\includegraphics[angle=0,width=90mm,  clip=false]{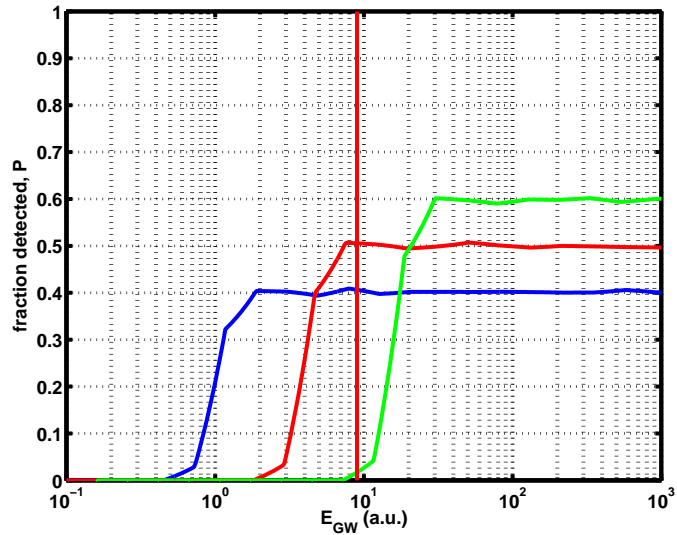} \label{fig:egwIndividual} }
\subfigure[]{
\includegraphics[angle=0,width=90mm,  clip=false]{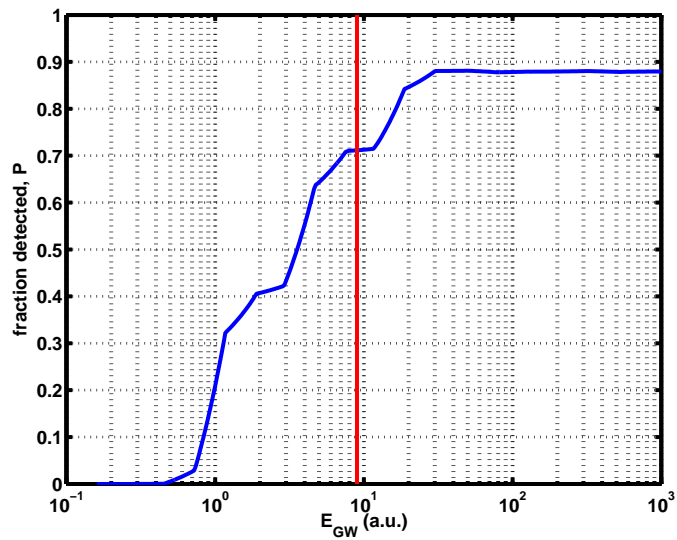} \label{fig:egwCombined} }
\caption{ (a) Individual GW emission energy ($\egw$) efficiency curves for the example shown in Figure\,\ref{fig:reachApproach}.  The red vertical line shows the actual $\egw$ predicted by the hypothetical model under examination.          
(b) Combined GW emission energy ($\egw$) efficiency curve for the example shown in Figure\,\ref{fig:reachApproach}.  The red vertical line at the predicted value $\egw=9$ intersects the curve at an exclusion confidence of 0.7, as before.} 
\end{center}
\end{figure}

Under this second approach, the individual efficiency curves look like Figure\,\ref{fig:egwIndividual}.  The combined exclusion confidence curve looks like Figure\,\ref{fig:egwCombined} and once again yields an exclusion confidence at nine units of energy of 0.7.  The first two CCSN should be detected with efficiencies given by the duty cycles, 0.4 and 0.5, while the third CCSN should not be detected and thus does not contribute to model exclusion. 

\section{Conclusion}

We have presented a method for estimating confidence levels for excluding theoretical models, given an observational result of null detections over multiple events.  Our method can potentially increase the exclusion confidence in cases where the detector duty cycle does not completely cover the signal regions.  This method could significantly improve the astrophysical potential of searches for gravitational waves from core collapse supernovae.

\ack
 
P.K. is grateful to the LIGO laboratory, supported by funding from United
States National Science Foundation. LIGO was constructed by the California Institute of Technology and
Massachusetts Institute of Technology with funding from
the National Science Foundation and operates under cooperative agreement PHY-0757058.
M.Z. would like to thank the COAS of ERAU for support, and the U.S. National Science Foundation for support through the grants NSF0855567 and NSF0919034.  S.K. acknowledges support by the U.S. National Science Foundation Grants No. PHY-1205512 and No. PHY-0855313 to the University of Florida, Gainesville, Florida. This paper has been assigned LIGO Document Number LIGO-\ligodoc.

\bibliographystyle{unsrt}

\section*{References}
\begin{thebibliography}{1}

\bibitem{detectorLivingReview}
Matthew Pitkin, Stuart Reid, , and James Hough.
\newblock Gravitational wave detection by interferometry (ground and space).
\newblock {\em Living Reviews in Relativity}, 14(5), 2011.

\bibitem{a5sgr}
J.~{Abadie} et~al.
\newblock {Search for Gravitational Wave Bursts from Six Magnetars}.
\newblock {\em \apjl}, 734(2):L35, 2011.

\bibitem{feldman98}
G.~J. {Feldman} and R.~D. {Cousins}.
\newblock {Unified approach to the classical statistical analysis of small
  signals}.
\newblock {\em Phys.~Rev.~D}, 57:3873--3889, April 1998.

\bibitem{S2S3S4GRB}
B.~{Abbott} et~al.
\newblock Search for gravitational waves associated with 39 gamma-ray bursts
  using data from the second, third, and fourth ligo runs.
\newblock {\em \prd}, 77(6):062004, 2008.

\bibitem{s5y1sgr}
B.~{Abbott} et~al.
\newblock Search for gravitational-wave bursts from soft gamma repeaters.
\newblock {\em \prl}, 101(21):211102, 2008.

\bibitem{stackSgr}
B.~P. {Abbott} et~al.
\newblock {Stacked Search for Gravitational Waves from the 2006 SGR 1900+14
  Storm}.
\newblock {\em \apjl}, 701:L68--L74, August 2009.

\bibitem{kalmus07}
P.~{Kalmus}, R.~{Khan}, L.~{Matone}, and S.~{M{\'a}rka}.
\newblock {Search method for unmodeled transient gravitational waves associated
  with SGR flares}.
\newblock {\em Classical and Quantum Gravity}, 24:659--+, October 2007.

\bibitem{kalmus08}
P.~{Kalmus}.
\newblock {\em Search for Gravitational Wave Bursts from Soft Gamma Repeaters}.
\newblock {PhD thesis}, Columbia University, 2008.
\newblock arXiv:0904.4394.

\end{thebibliography}

\end{document}